\documentclass[runningheads]{llncs}
\usepackage[T1]{fontenc}
\usepackage{graphicx}
\usepackage[nolist]{acronym}
\usepackage{xspace}
\usepackage{todonotes}
\usepackage{hyperref}

\usepackage{subcaption}

\newcommand{\SystemName}{\textsc{FedBlockParadox}\xspace}
\newcommand{\systemName}{FedBlockParadox\xspace}
\newcommand{\mypar}[1]{\smallskip\noindent\textbf{#1.}}

\begin{document}

\begin{acronym}
    \acro{PoW}{Proof of Work}
    \acro{PoF}{Proof of Federation}
    \acro{IID}[iid]{independent and identically distributed}
\end{acronym}

\title{Poster: \SystemName  \xspace- A Framework for Simulating and Securing Decentralized Federated Learning}
\subtitle{\normalfont\scriptsize This poster paper has been accepted at the International Conference on Detection of Intrusions and Malware, and Vulnerability Assessment (DIMVA '25)}
\titlerunning{Poster - \SystemName}
%
\author{
Gabriele Digregorio \and
Francesco Bleggi \and
Federico Caroli \and
Michele Carminati \and
Stefano Zanero \and
Stefano Longari
}

\authorrunning{G. Digregorio et al.}
%
\institute
{NECSTLab, DEIB, Politecnico di Milano, Milano, Italy \\
\email{\{gabriele.digregorio, michele.carminati, stefano.zanero, stefano.longari\}@polimi.it} \\
\email{\{francesco.bleggi, federico.caroli\}@mail.polimi.it}}
\maketitle              
\begin{abstract}
A significant body of research in decentralized federated learning focuses on combining the privacy-preserving properties of federated learning with the resilience and transparency offered by blockchain-based systems. While these approaches are promising, they often lack flexible tools to evaluate system robustness under adversarial conditions. To fill this gap, we present \systemName, a modular framework for modeling and evaluating decentralized federated learning systems built on blockchain technologies, with a focus on resilience against a broad spectrum of adversarial attack scenarios. It supports multiple consensus protocols, validation methods, aggregation strategies, and configurable attack models. By enabling controlled experiments, \systemName provides a valuable resource for researchers developing secure, decentralized learning solutions. The framework is open-source and built to be extensible by the community.

\end{abstract}

\section{Introduction}
In the rapidly evolving landscape of decentralized systems, federated learning holds considerable potential for enhancing privacy and scalability across various applications, fundamentally transforming data management and usage in diverse sectors such as healthcare, finance, and vehicular systems~\cite{10880750,10770536,digregorio2024evaluating}. However, its integration into real-world scenarios introduces substantial challenges. Traditional federated learning frameworks, which rely on a central coordinating authority, are inherently vulnerable to malicious attacks and present single points of failure~\cite{bouacida2021vulnerabilities}. This central dependency not only increases the likelihood of security breaches but also creates potential bottlenecks in data processing and might introduce bias during model aggregation.
To address these limitations, recent advancements have explored the elimination of the central orchestrator by developing fully decentralized federated learning systems using blockchain technology~\cite{li2020blockchain}. However, while these approaches mitigate certain risks, they also introduce complexities in managing and validating distributed model updates, as well as in accurately assigning the various roles necessary to support the system. These aspects are often overlooked in the current state of the art, with many solutions lacking robust guarantees against sophisticated adversarial attacks. Moreover, the absence of consistent experimental settings makes it difficult to perform meaningful comparisons across architectures.
To enable the evaluation, validation, and comparison of different decentralized federated learning systems, we introduce \systemName, a framework for simulating complex decentralized federated learning environments in a customizable manner. \systemName supports key features of decentralized federated learning, including various consensus algorithms, validation mechanisms, and heterogeneous data distributions. It is designed to help the research community assess the ability of decentralized systems to withstand known adversarial attacks.

The main contributions of our work are as follows:
\begin{itemize}
    \item We present \systemName, a framework for evaluating the resilience of decentralized federated learning approaches, particularly under adversarial conditions often overlooked in existing implementations.
    \item We make \systemName highly configurable, supporting a range of validation mechanisms, aggregation techniques, and consensus algorithms. This flexibility enables testing across diverse and realistic scenarios. While the framework includes several integrated algorithms by default, it is intended to be extensible, allowing the community to incorporate and evaluate new, more sophisticated approaches under consistent experimental settings to enable fair comparisons.
    \item We release \systemName as open-source software, freely available to the community~\footnote{\label{footnote}\url{https://github.com/necst/FedBlockParadox}}. Our goal is to support the empirical validation of both existing and future solutions that combine blockchain technology with federated learning.
\end{itemize}
\section{State of the Art}

We review key state-of-the-art proposals for decentralized federated learning combined with blockchain technology. Interested readers may refer to additional works summarized in the surveys presented in~\cite{qu2022blockchain,moore2023survey,zhu2023blockchain,nguyen2021federated}.

In~\cite{ma2022federated}, the authors present an implementation of decentralized federated learning that incorporates the \ac{PoW} consensus mechanism.
However, this approach does not scale well in networks with large numbers of nodes, as each node—regardless of its computational capability—must download all updates from the previous round and perform local aggregation. Furthermore, the system lacks a validation mechanism to detect and discard malicious updates.

The authors of~\cite{li2020blockchain} integrate blockchain technology with a novel committee-based consensus mechanism. In each round, a dynamic subset of nodes, referred to as the \textit{committee}, is elected to perform validation. This subset evaluates submitted model updates using local datasets and assigns scores based on validation accuracy. Only updates meeting predefined accuracy thresholds are aggregated into the global model. 
However, the paper does not explore the effectiveness of alternative validation mechanisms beyond accuracy-based filtering.

In~\cite{qu2020decentralized}, the authors propose an implementation of decentralized federated learning tailored for fog computing environments. The architecture integrates local updates from end devices into a distributed ledger, verified through a \ac{PoW} consensus mechanism. A central challenge discussed is the trade-off between privacy protection and system efficiency. Nevertheless, similar to~\cite{ma2022federated}, this approach struggles to scale in large networks, as every node must download and locally aggregate all model updates from each round.

The work in~\cite{zhou2023decentralized} introduces a decentralized federated learning architecture that leverages node selection and knowledge distillation to overcome limitations of traditional centralized federated learning. A dynamically selected central node replaces the static server, based on individual node performance and data quality. However, the architecture lacks defenses against malicious behavior such as targeted poisoning or falsified dataset sizes. Moreover, no mechanism is provided to regulate or verify the actions of the elected central node, allowing it to potentially bias the global model without detection.

In~\cite{mao2024blockchain}, a set of nodes replaces the traditional central server, coordinating model updates while preserving system integrity. The aggregation mechanism dynamically adapts to the state of participant data and models by weighting updates based on validation loss and training data volume. A validation step based on accuracy is employed to filter out low-quality or malicious updates. Malicious participants submitting poisoned updates are penalized accordingly.

The authors of~\cite{shayan2020biscotti} introduce a novel consensus protocol termed \ac{PoF}, which assigns roles to peers based on their earned stake, accumulated through valid contributions to the global model. This role distribution promotes resilience against centralized control and reduces the likelihood of collusion. The authors also incorporate the Multi-Krum algorithm to detect and reject anomalous updates that deviate significantly from the majority, thereby enhancing robustness against poisoning attacks.
\section{\systemName} 
We introduce \systemName, a framework that simulates decentralized federated learning systems under adversarial conditions. Its ultimate goal is to provide a reliable means to test and validate various defenses against common attacks in decentralized federated learning.
To support this objective, we release our framework as an open-source tool, encouraging further research and testing of proposed solutions. \systemName allows customization of key parameters, including consensus algorithms, validation mechanisms, aggregation strategies, and update-sharing methods. 
The configuration options available at the time of writing are inspired by state-of-the-art implementations in decentralized federated learning systems and detailed below. \systemName is designed to be extensible,  encouraging the integration of additional methods.





\mypar{General Configurations}
These settings define the core parameters of \systemName, including the total number of simulation rounds, the model architecture and initialization, and the number of participating nodes. Moreover, \systemName supports model updates in both weight and gradient form.

\mypar{Consensus Algorithms}
\systemName currently supports Proof-of-Work (PoW), Proof-of-Stake (PoS), and committee-based consensus, each supporting further customization within the framework.

\mypar{Validation Algorithms}
\systemName supports a range of validation algorithms. It includes naive strategies such as Pass-Weights and Pass-Gradients, which accept all updates without verification. More advanced methods, such as Global and Local Dataset Validation, evaluate model updates against a shared or local dataset, accepting them only if they meet a predefined accuracy threshold. Finally, Multi-Krum Validation allows validators to assess updates collectively by measuring pairwise distances and filtering out the most divergent ones.

\mypar{Aggregation Algorithms}
\systemName currently supports three aggregation strategies: FedAvg~\cite{mcmahan2017communication} for model updates in weight form, and Mean and Median Aggregation for model updates in both gradient and weight form. 

\mypar{Malicious Nodes and Attacks}
\systemName allows specific nodes to exhibit malicious behavior. Supported attack types include label flipping, targeted data poisoning, and additive noise. Each attack type is configurable to suit the simulation context and can be tailored to evaluate system robustness under different adversarial scenarios.

\mypar{Dataset Settings}
The simulator supports the use of different datasets. In our initial implementation, we employed the MNIST~\cite{lecun1998mnist} and CIFAR-10~\cite{krizhevsky2009learning} datasets. Key configuration options include the number of dataset partitions, the percentage of \ac{IID} versus non-\ac{IID} partitions, and the strategy for assigning partitions to nodes. 


\section{Preliminary Experiments}
\begin{figure}[tb]
    \centering
    \makebox[\textwidth][c]{%
        \begin{subfigure}{0.45\textwidth}
            \centering
            \includegraphics[width=\textwidth]{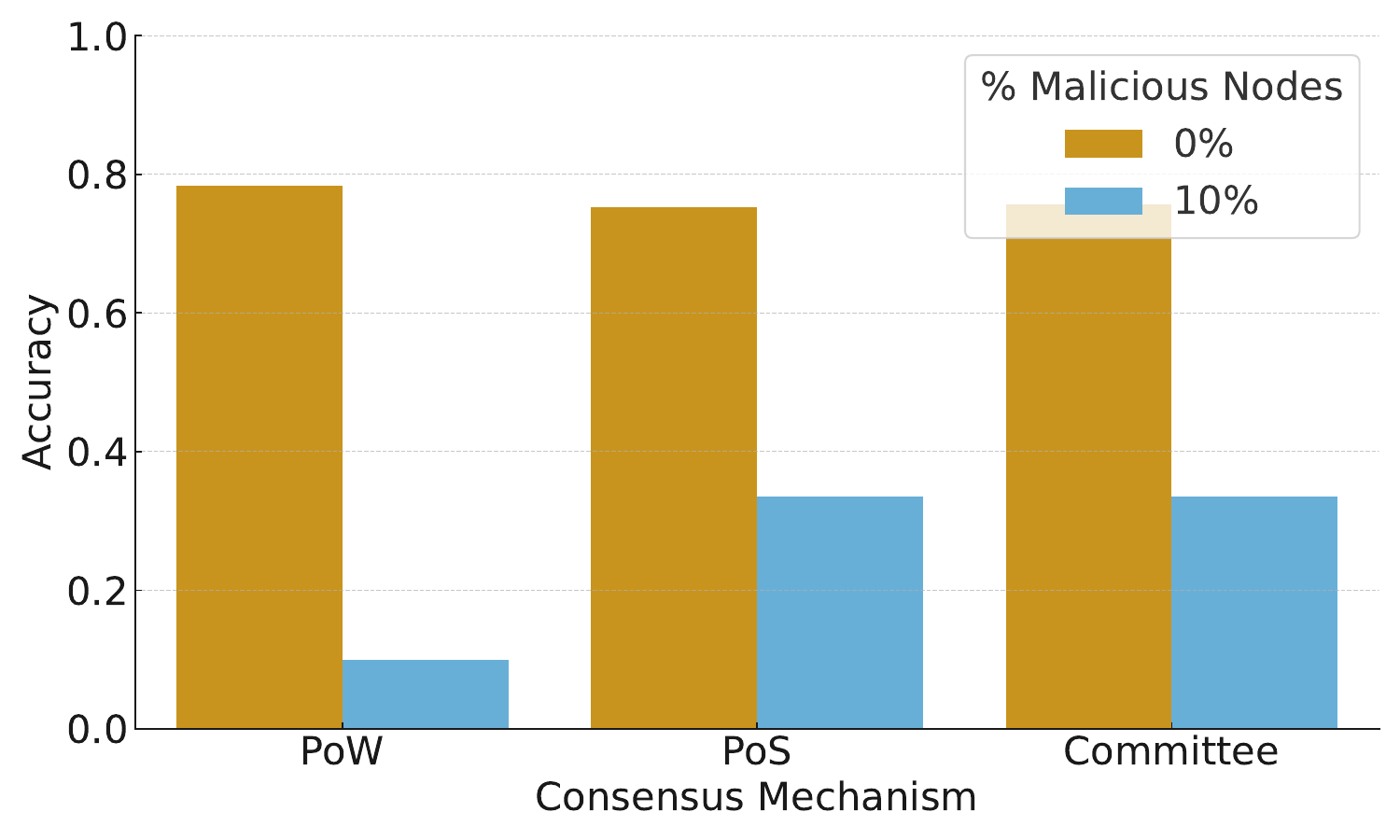}
            \caption{Additive noise attack results.}
            \label{fig:4gb}
        \end{subfigure}
        \hfill
        \begin{subfigure}{0.45\textwidth}
            \centering
            \includegraphics[width=\textwidth]{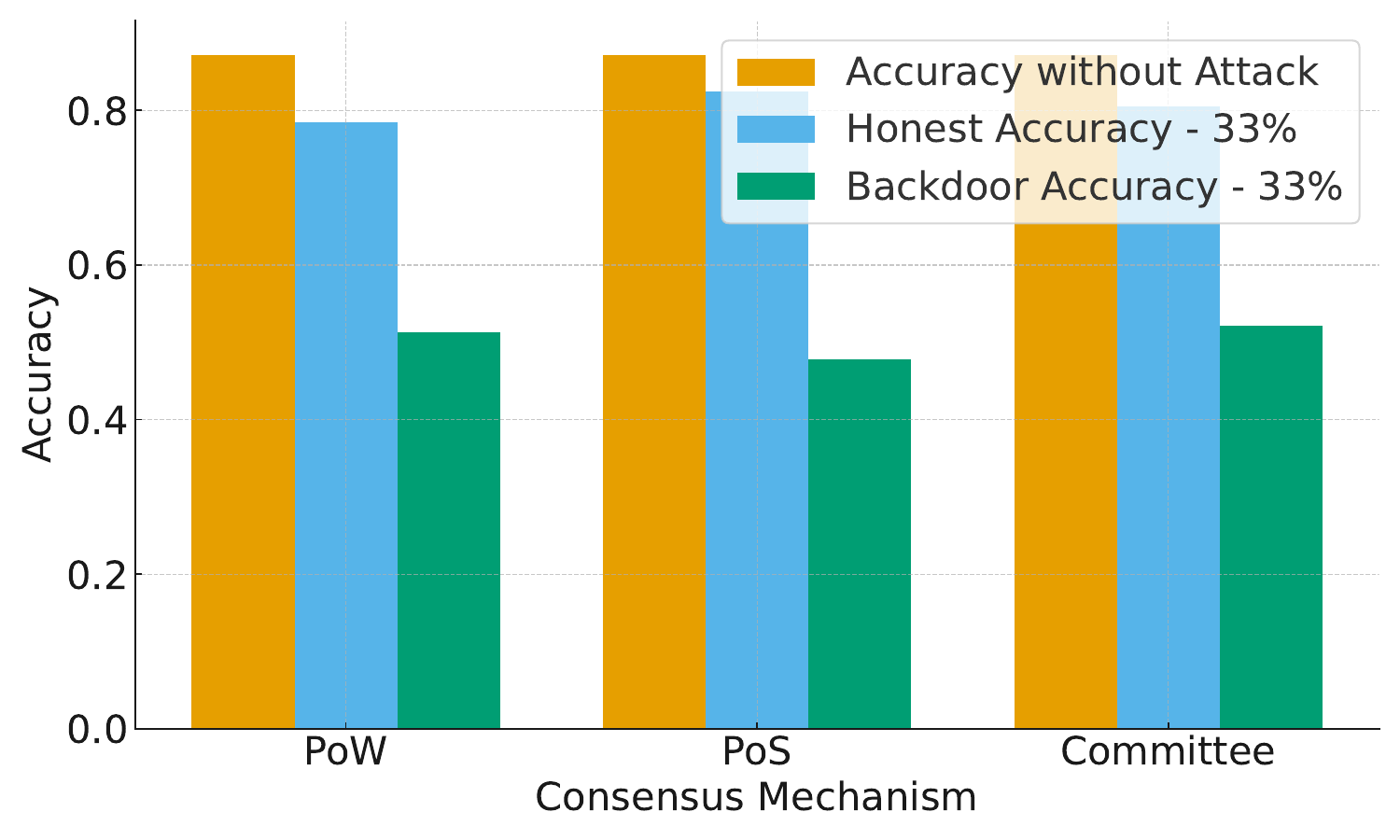}
            \caption{Data poisoning attack results.}
            \label{fig:16gb}
        \end{subfigure}
        \hfill
        \begin{subfigure}{0.45\textwidth}
            \centering
            \includegraphics[width=\textwidth]{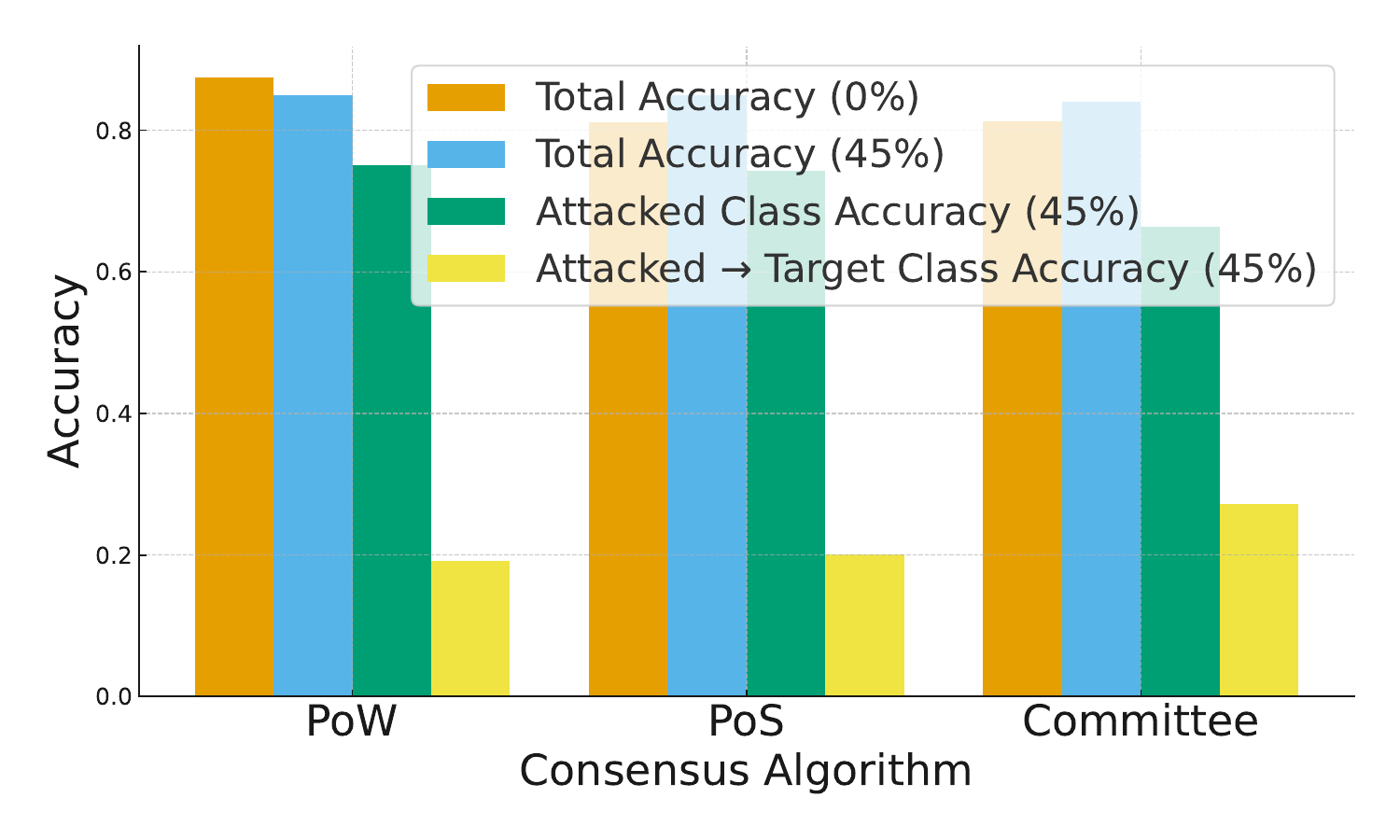}
            \caption{Label flipping attack results.}
            \label{fig:32gb}
        \end{subfigure}
    }
    \caption{Aggregated preliminary results of attacks on the datasets with various parameters, e.g., attacker amounts and aggregation methods.}
    \label{fig:preliminary}
\end{figure}

We ran a set of experiments to validate our tool and provide baseline performance metrics. The full list of experiments is available in our repository\textsuperscript{\ref{footnote}}, we briefly sum up the results for each attack category in Figure~\ref{fig:preliminary}. Tests have been developed varying aggregation methods, validation processes, datasets, architecture, consensus algorithms, and amount of malicious nodes.

\section{Conclusion}
 In this work, we presented \systemName, a simulation framework tailored for modeling decentralized federated learning environments using blockchain-based architectures. Through its modular design and configurability, \systemName enables experimentation with consensus protocols, validation mechanisms, aggregation techniques, and attack strategies. Our framework addresses critical gaps in the current state of the art by supporting reproducible testing and community-driven extensibility. By releasing \systemName as open-source software, we encourage the development of new defense mechanisms and validation approaches to advance the field of secure decentralized federated learning. 

 \mypar{Future Work} As a research-enabling framework, FedBlockParadox will be employed to empirically and systematically evaluate novel defense strategies and test security assumptions in blockchain-based decentralized federated learning systems proposed in the current state of the art. By providing consistent experimental settings and a standardized evaluation pipeline, it enables meaningful comparisons across architectures and approaches. Our ultimate goal is to assess the real-world impact of blockchain-enabled decentralization on the robustness and security of these systems.

\section*{Acknowledgements}
This study was carried out within the MICS (Made in Italy – Circular and Sustainable) Extended Partnership and received funding from Next-Generation EU (Italian PNRR – M4 C2, Invest 1.3 – D.D. 1551.11-10-2022, PE00000004). CUP MICS D43C22003120001.

%
%
%
%
\bibliographystyle{splncs04}
\bibliography{biblio}

\begin{thebibliography}{10}
\providecommand{\url}[1]{\texttt{#1}}
\providecommand{\urlprefix}{URL }
\providecommand{\doi}[1]{https://doi.org/#1}

\bibitem{lecun1998mnist}
The mnist database of handwritten digits. http://yann. lecun. com/exdb/mnist/

\bibitem{10770536}
Boiano, A., Di~Gennaro, M., Barbieri, L., Carminati, M., Nicoli, M., Redondi, A., Milasheuski, U., Kianoush, S., Savazzi, S., Aillet, A.S., Santos, D.R., Serio, L.: A secure and trustworthy network architecture for federated learning healthcare applications. In: 2024 20th International Conference on Wireless and Mobile Computing, Networking and Communications (WiMob). pp. 124--129 (2024). \doi{10.1109/WiMob61911.2024.10770536}

\bibitem{bouacida2021vulnerabilities}
Bouacida, N., Mohapatra, P.: Vulnerabilities in federated learning. IEEe Access  \textbf{9},  63229--63249 (2021)

\bibitem{digregorio2024evaluating}
Digregorio, G., Cainazzo, E., Longari, S., Carminati, M., Zanero, S.: Evaluating the impact of privacy-preserving federated learning on can intrusion detection. In: 2024 IEEE 99th Vehicular Technology Conference (VTC2024-Spring). pp.~1--7. IEEE (2024)

\bibitem{krizhevsky2009learning}
Krizhevsky, A., Hinton, G., et~al.: Learning multiple layers of features from tiny images  (2009)

\bibitem{li2020blockchain}
Li, Y., Chen, C., Liu, N., Huang, H., Zheng, Z., Yan, Q.: A blockchain-based decentralized federated learning framework with committee consensus. IEEE Network  \textbf{35}(1),  234--241 (2020)

\bibitem{ma2022federated}
Ma, C., Li, J., Shi, L., Ding, M., Wang, T., Han, Z., Poor, H.V.: When federated learning meets blockchain: A new distributed learning paradigm. IEEE Computational Intelligence Magazine  \textbf{17}(3),  26--33 (2022)

\bibitem{mao2024blockchain}
Mao, Q., Wang, L., Long, Y., Han, L., Wang, Z., Chen, K.: A blockchain-based framework for federated learning with privacy preservation in power load forecasting. Knowledge-Based Systems  \textbf{284},  111338 (2024)

\bibitem{mcmahan2017communication}
McMahan, B., Moore, E., Ramage, D., Hampson, S., y~Arcas, B.A.: Communication-efficient learning of deep networks from decentralized data. In: Artificial intelligence and statistics. pp. 1273--1282. PMLR (2017)

\bibitem{moore2023survey}
Moore, E., Imteaj, A., Rezapour, S., Amini, M.H.: A survey on secure and private federated learning using blockchain: Theory and application in resource-constrained computing. IEEE Internet of Things Journal  \textbf{10}(24),  21942--21958 (2023)

\bibitem{nguyen2021federated}
Nguyen, D.C., Ding, M., Pham, Q.V., Pathirana, P.N., Le, L.B., Seneviratne, A., Li, J., Niyato, D., Poor, H.V.: Federated learning meets blockchain in edge computing: Opportunities and challenges. IEEE Internet of Things Journal  \textbf{8}(16),  12806--12825 (2021)

\bibitem{qu2020decentralized}
Qu, Y., Gao, L., Luan, T.H., Xiang, Y., Yu, S., Li, B., Zheng, G.: Decentralized privacy using blockchain-enabled federated learning in fog computing. IEEE Internet of Things Journal  \textbf{7}(6),  5171--5183 (2020)

\bibitem{qu2022blockchain}
Qu, Y., Uddin, M.P., Gan, C., Xiang, Y., Gao, L., Yearwood, J.: Blockchain-enabled federated learning: A survey. ACM Computing Surveys  \textbf{55}(4),  1--35 (2022)

\bibitem{10880750}
Santos, D.R., Aillet, A.S., Boiano, A., Milasheuski, U., Giusti, L., Di~Gennaro, M., Kianoush, S., Barbieri, L., Nicoli, M., Carminati, M., Redondi, A.E.C., Savazzi, S., Serio, L.: A federated learning platform as a service for advancing stroke management in european clinical centers. In: 2024 IEEE International Conference on E-health Networking, Application \& Services (HealthCom). pp.~1--7 (2024). \doi{10.1109/HealthCom60970.2024.10880750}

\bibitem{shayan2020biscotti}
Shayan, M., Fung, C., Yoon, C.J., Beschastnikh, I.: Biscotti: A blockchain system for private and secure federated learning. IEEE Transactions on Parallel and Distributed Systems  \textbf{32}(7),  1513--1525 (2020)

\bibitem{zhou2023decentralized}
Zhou, Z., Sun, F., Chen, X., Zhang, D., Han, T., Lan, P.: A decentralized federated learning based on node selection and knowledge distillation. Mathematics  \textbf{11}(14), ~3162 (2023)

\bibitem{zhu2023blockchain}
Zhu, J., Cao, J., Saxena, D., Jiang, S., Ferradi, H.: Blockchain-empowered federated learning: Challenges, solutions, and future directions. ACM Computing Surveys  \textbf{55}(11),  1--31 (2023)

\end{thebibliography}

\end{document}